# THE ROLE OF GAME JAMS IN DEVELOPING INFORMAL LEARNING OF COMPUTATIONAL THINKING: A CROSS-EUROPEAN CASE STUDY

## H. Boulton[1], B. Spieler[2], A. Petri[2], C. Schindler[2], W. Slany[2], X. Beltran[3]

[1] Nottingham Trent University (UNITED KINGDOM)
[2] Graz University of Technology (AUSTRIA)
[3] Inmark (SPAIN)

This paper will present a cross-European experience of game jams as part of a Horizon 2020 funded project: No-one Left Behind (NOLB). The NOLB project was created to unlock inclusive gaming creation and experiences in formal learning situations from primary to secondary level, particularly for children at risk of social exclusion. The project has engendered the concept of game jams, events organised with the aim of designing and creating small games in a short time-frame around a central theme. Game jams can support engagement with informal learning beyond schools across a range of disciplines, resulting in an exciting experience associated with strong, positive emotions which can significantly support learning goals. This paper will disseminate experience of two cross-European game jams; the first a pilot and the second having over 95 submissions from countries across Europe, America, Canada, Egypt, the Philippians and India. Data collected through these games jams supports that coding, designing, reflection, analysing, creating, debugging, persevering and application, as well as developing computational thinking concepts such as decomposition, using patterns, abstraction and evaluation. The notion of game jams provides a paradigm for creating both formal and informal learning experiences such as directed learning experience, problem-solving, hands-on projects, working collaboratively, and creative invention, within a learner-centred learning environment where children are creators of their own knowledge and learning material. This paper explores the use of a mobile app, Pocket Code, in schools across Europe in two game jams during the academic year 2015-16 with children aged 11-18. Pocket Code provides an environment which supports learners in easily creating apps directly on their smart-phones and tablets through a visual Lego®-style programming language where users can put code bricks together to form scripts. We draw on a range of data to support how game jams can be used as a design research method to observe the creation of knowledge in fast-paced, collaborative environments across a range of disciplines. Our data evidences that learners can be more motivated through game jams and that learners who are less likely to create games are nevertheless more engaged in a game jam setting. We will also present the frameworks for 3 games from different disciplines: Chemistry, Languages, and Mathematics.

## 1. INTRODUCTION

With the increased emphasis on computational thinking in curricula across Europe schools are identifying different learning experiences for learners which will engage them in computing. For example in the United Kingdom (UK) the Government dis-established the Information Communications Technology curriculum, replacing it in 2014 with a new Computing Curriculum. This curriculum is regulatory from primary (5-11 years) through secondary school (11-16). Learners wishing to take qualifications in computing can do so through General Certificate of Secondary Education examination (age 16) or through AS and A2 examination (16-18 years). Schools across Europe are increasingly adapting game jams as a fun method of engaging learners in learning computing and developing their computational skills.

Within gaming environments, learners are encouraged to be creative and collaborate with others, working either in groups or individually. The internet enables competition to develop across countries, and software developments enable cross-platform gaming on the same theme (Chandrasekaran, Stojcevski, Littlefair, & Joordens, 2012). This paper reports how game jams have been used to inspire engagement in using game jams in learning in different disciplines across three countries as part of a Horizon 2020 funded project, No-one Left Behind (NOLB)[1].

The NOLB project was created to unlock inclusive gaming creation and experiences in formal and informal learning situations from primary to secondary level, particularly for children at risk of social exclusion such as those with special educational needs and disabilities, immigrants, and also recognising gender differences in engaging with computing and game jams. The project is informing the development of a new generation of Pocket Code (a mobile media-rich programming environment for children) to unlock inclusive gaming creation and experiences in formal and informal learning situations, providing meaningful learning experiences and supporting learners to realise their full potential; by transferring game mechanics, dynamics, assets and in-game analytics from non-leisure digital games into Pocket Code, which also will be adapted to academic curricula.

The project based in the UK, Austria and Spain, spans five different curriculum disciplines including humanities, science. Partners within each country are working with a different type of learner: Austria is focussing improving girls' interest in STEM subjects (science, technology, engineering, and mathematics) through fostering social inclusion in class communities; Spain is focussing on learners who are at risk of social exclusion related to immigration through fostering collaboration, cooperation, and engagement between Spanish and immigrant pupils; the UK team are working with children who have special educational needs and disabilities.

The NOLB project encourages learners to use a development environment to create games directly on their mobile devices to enhance their abilities across all academic subjects, including logical reasoning, creativity, and develop social skills. Participants in traditional game jams are typically experienced male developers (Kafai, 2006); given the nature of the NOLB project there was a deliberate intention to appeal more widely to girls, with mixed success as discussed below. The NOLB project has identified four game genres applicable to the game jams: quiz games, adventure games, puzzle games, and action games.

The NOLB project has engendered the concept of game jams, events organised with the aim of designing and creating small games in a short time-frame around a central theme. Game jams can support engagement with informal learning beyond schools across a range of disciplines, resulting in an exciting experience associated with strong, positive emotions which can significantly support learning goals. Learning is complex and in general has to be guided in order not to waste time and energy or lead to misconceptions or incomplete knowledge (Razak, Connolly, & Hainey, 2011). This can be a challenging task when introducing playful elements within the school curricula (Petri, Schindler, Slany, Spieler, & Smith, 2015) such as game jams. Throughout the duration of the NOLB project game jam events will be hosted, using Pocket Code, across Europe such as the Global Game Jam, i<tag> Hackaton, and Computer Science Education Week, in the form of Pocket Code Game Jams. This paper will disseminate experience of two cross-European Game Jams; the first a pilot and the second having over 95 submissions from countries across Europe, America, Canada, Egypt, the Philippians and India. Data collected through these game jams supports that coding, designing, reflection, analysing, creating, debugging, persevering and application, as well as developing computational thinking concepts such as decomposition, using patterns, abstraction and evaluation. Game jams can contribute to improve learner's overall motivation and productivity (Domínguez, *et al,* 2013) but teachers need to be supported and engaged as well. This paper presents early game jam experiences with pupils from 11 to 18 years across two events. The paper sets out to introduce the reader to the mobile game development tool used in this project, Pocket Code and sets out experiences of both game jams.

Game jams can foster collaborative working through small teams working with distributed tasks, to develop games reflecting creativity and competition among learners, while learning computing concepts (Chatham, *et al*, 2013). Game jams integrate various game making disciplines, such as programming, art and design, and literacy, to create games under given constraints such as a fixed time frame or topic. The goal of the game jams reported in this paper was to focus on networking, collaboration, engage learners and provide a positive first contact with coding tools. Within game jams certain rules can be adopted to constrain the design space and scope (Deen *et al*, 2014). A key element of the game jams was creating games within a specific theme within a given period of time (Kaitila, 2012). The learners involved were encouraged to work quickly, be creative in their designs, work collaboratively, and finish their game within a given deadline. Learners were provided with initial goals, followed by additional sub-goals within the gaming periods [2]. For both game jams discussed in this paper the themes were chosen to reflect topics situated around subject disciplines (Goddard, Byrne, & Mueller, 2014) where factors such as learning achievement, engagement, and persistence, as well as the development of computing and computational thinking skills are important. In each game jam schools involved were encouraged to facilitate the presentation of the games to a wider audience such as a whole class or year group where possible, in order to provide positive feedback and recognition to the students that have put extra effort. Thus acknowledgement by teachers and peers, and knowing of achievement was aimed to be seen, appreciated and celebrated.

The NOLB game-making framework is comprises the concepts, content and methodologies to transfer into practice the processes (curriculum adaptation/planning, teaching-learning processes, and assessment and feedback) that link the three pillars of the NOLB conceptual framework (see figure 1). The integration of these three processes into the NOLB game making framework supported the implementation of the New Generation of Pocket Code in the game jams.

Pocket Code is an environment which supports learners in easily creating games directly on their smartphones and tablets 3. Modern smartphones are equipped with a large number of sensors, although most mobile games only use few or none of them (Kafai and Vasudevan, 2015). Within Pocket Code one can create games using the device's sensors such as inclination, acceleration, loudness, face detection or the compass direction, which makes user input easy and engaging. The feature to merge programs among users and transfer objects, code, looks, and sounds between projects via the "backpack" functionality, fosters distributed development and collaboration among learners. With Pocket Code it is also possible to connect via Bluetooth to Lego Mindstorms ® robots or ArduinoTM boards. These computational construction kits make creating programmable hardware accessible to even novice designers and combines coding and crafting with a rich context for engaging learners (Kafai, Searle and Fields, 2014). Pocket Code has a visual Lego® -style programming language using bricks to form scripts, thus supporting the development of games. Learners who have used similar programmes, such as Scratch[4] can easily transfer this learning to Pocket Code which has a similar visual interaction style and elements. Learners who have programming experience with Scratch5, a popular programming platform in the UK, are able to immediately create programs with Pocket Code's similar visual interaction style and elements. There are several key advantages to using Pocket Code for game jams. For example Pocket Code has inbuilt sensors, such as inclination, acceleration, and loudness, which are easy to utilise in games and add additional elements to game design used in game jams. In addition learners can merge programs among users and transfer objects, code, looks, and sounds between projects via a 'Backpack'' functionality which in turn fosters distributed development.

Pocket Code is ideal for creating games for game jams within a short time span in fast paced and collaborative environments. The collaborative nature of game jams enable learners to work together while expressing their individual ideas and creativity (Chatham, Pijnappel, & Mueller, 2013). Game jams integrate various game making disciplines such as programming, art and design, literacy, and creativity, to create games under constraints such as a fixed time frame or topic. The goal of a game jam is to focus on networking, collaboration and, most importantly, a positive first contact with coding tools.

NOLB goals for exploring the use of game jams as an integral element to this project are to:

- explore Game Jams as a research method;
- identify the advantages of this method;
- identify problems such as difficulties in generalizing results and missing functionality in Pocket Code;
- introduce students to competitive approaches to increase motivation;
- evaluate participation and acknowledgement of participation as a reward mechanism;
- increase the game making skills in limited and short times to support easy development and participation in classes;
- Inspire meaningful learning by providing a unique interpretation and direction of academic content through personal game rules and selected game genre;
- develop playable games as example projects;
- create opportunities for official online Game Jams with Pocket Code.

---

## 2. THEORETICAL FRAMEWORK

The theoretical framework for this study is based on the theory of constructionism, which emphasizes design and sharing of artefacts (Parmaxi, & Zaphiris, 2014). Jonassen (1999) describes how in creating a constructivist learning environment, a teacher nurtures learning of concepts and problem solving within computational thinking. We have chosen the goal-based scenario (GBS) methodology as our rationale to match cognitive skills, reflecting a constructionist approach to learning. Learners are encouraged to construct games, making the design decisions themselves, thus developing new relationships with knowledge, rather than embedding lessons within games (Kafai, 2006). The NOLB project examines how to attract, motivate, and engage learners with content from different academic discipline curricula and at the same time supporting the learning process and providing an effective and dynamic learning experience.

Through the use of Pocket Code the project aims to become an empowering tool that supports the constructionist approach and therefore the development of creativity, problem solving, logical thinking, system design, and collaboration skills.

The NOLB project has developed a game making framework which interlinks the use of Pocket Code within a games environment into the classroom through the development of games within subject disciplines. Within the wider elements of the NOLB project, that is beyond game jams, the development of games within subject disciplines will be related to classroom pedagogy and include direct links to curriculum planning resulting in assessment and feedback. Ultimately findings and experiences from the game jams reported in this paper will feed into a new generation of Pocket Code.

## 3. METHODS

### 3.1 Data Collection

The following data were collected:

- Raw data from Facebook, Twitter and the game jam website;
- Observations;
- Number of submissions and details of those submitting such as country, gender and age;
- Questionnaire. This was sent to all learners who had submitted a game. Both qualitative and quantitative data were collected – this included demographic data, programming language used, motivations to take part, the theme, amount of time spent on their game, whether they worked alone or with a team, size of team where relevant, place where the game was developed (school, home, in transit), diversifiers they had utilised, observations of the time frame, and how they had found out about the game jam.

The raw data collected from social media is set out in the table below:

| AliceGameJam.com | |
|---|---|
| Sessions: | 3442 |
| Pages/Session: | 2.37 |
| Bounce-Rate | 37.8% |
| Avg. Session Duration | 00:02 |
| Facebook | |
| Likes: | 498 |
| Reach Promo English, eg Australia, South Africa, Poland | 9821 |
| Reach Promo Spanish, eg Spain and South America | 35978 |
| Reach Promo Total: | 45799 |
| Target Group (12-17 fem) | 83% |
| Twitter | |
| Impressions: | 57500 |
| Link-Clicks: | 37 |

| Interactions % | | 0,8% | |
|---|---|---|---|

Table 1: Social media data

Data were collected which suggested interest from over 100 countries either through a game submission, taking part in a website activity, or involvement with the game jam through social media.

The total number of submissions by country is shown below:

| Country | Submissions | Pocket Code | Scratch |
|---|---|---|---|
| Austria | 16 | 16 | |
| Bosnia Her. | 1 | 1 | |
| Canada | 1 | 1 | |
| Egypt | 1 | 1 | |
| Germany | 1 | 1 | |
| Hungary | 1 | | 1 |
| India | 20 | | 20 |
| Italy | 31 | 6 | 25 |
| Philippines | 1 | 1 | |
| Spain | 4 | 4 | |
| United Kingdom | 8 | 8 | |
| United States | 3 | | |
| Other | 18 | | |
| Total | 106 | | |

Table 2: Programme code per country

The data indicated that of those learners who completed the questionnaire 31 had developed their game at school, 59 had developed it at home and 5 had developed it in transit.

## 3.2 Game Jams - Research tool & method

NOLB is transferring game mechanics and dynamics to classes by adapting Game Jams. The project is facilitating changes in learners' academic engagement and acquisition of knowledge by introducing Game Jam elements. Through the game jams we are enabling learners to take control of their learning and expose them to different ways of exploring academic content leading to greater engagement, persistence, and potentially a deeper level of academic knowledge. The Pocket Code software is already able to produce some analytics relating to learner engagement.  As the project develops further and the outcomes of the game jams and use of Pocket Code in classrooms are analysed we will further develop the software to gather additional analytics which will support teachers in curriculum planning.

For the NOLB project the game jam approach bridges the gap between gaming-techniques and pedagogy across academic subjects, grades, schools, and national boundaries (Goddard, Byrne, & Mueller, 2014). The main aims of our approach are:

- to bring learners from different contexts and levels of knowledge together;
- to motivate them through blended interplay experiences to create new knowledge structures;
- to assess the potential of playful, experiential, and participatory learning experience;
- to devise a course structure that facilitates interdisciplinary and multinational encounters;
- to evaluate the competences and benefits of providing innovative perspectives;
- To provide feedback of integration of game methodologies in learning experiences to improve further academic planning and teaching delivery.

Through the game jam concept different and novel learning opportunities will be facilitated enabling learning to transfer from the classroom in a school to national and international levels. One of the key elements of this project is the integration of concepts used in commercial games to provide a leverage the power of games through providing engaging, goal-oriented interactive experiences which make recognizable connections between leisure activities which excite young game players on their academic endeavours. We have been able to draw upon the expertise of professional game-makers to inspire the creative efforts of a new generation of digital makers.

## 3.3 Settings for Pocket Code Game Jam

To encourage schools to use Pocket Code the project team organised online game jams through the Pocket Code Web interface and an optional landing page that contains all information about the event. Via a web interface, tutorials, graphics, and other assets are provided and the game submission process is managed. In the future we plan to let teachers create their own game jam events using this platform. We provided:

- A general theme (announced ahead of the event).
- A surprise theme (announced at the start of the event).
- Diversifiers to increase engagement and spark creativity in the development process.
- Topic related support material (graphics, artefacts, and tutorials).
- Promotion material for schools.
- A submission website and final questionnaire.

Additional aspects included:

- Working in small groups (two to three pupils).
- "The Shape of a Game" ceremony: Learners are encouraged to use good design principles[6]. This framework integrates screens required for game design and is structured as follows: A Title Screen, an Instructions Screen (explains both, the "goal" and "rules"), one or more Levels and a Game End Screen.
- Awareness of license issues and attribution.

After submission, participants were asked to give feedback via questionnaires. This feedback is analysed to identify improvements of Pocket Code and the game jam methodology.

The framework for our game jams is shown below:

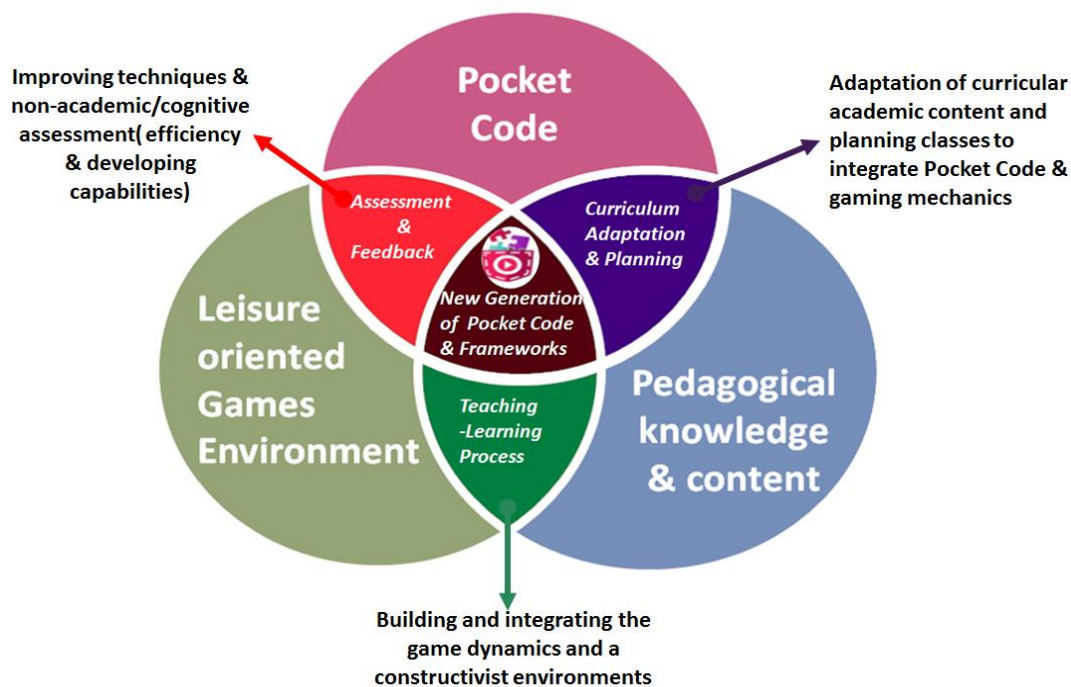

*Figure 1: NOLB game jam framework*

## 4. DISCUSSION

### 4.1 Pilot

Prior to the pilot study every pilot site conducted a feasibility study at selected schools who were to take part in the game jam. As this was the first time Pocket Code was utilised for the purpose of a game jam instruction was provided for teachers to learn how to use Pocket Code. Feedback was collected through observation during classes and questionnaires at the beginning and the end of the feasibility study.

A first test run of the game jam was conducted during the European Code Week in October 2015 with first year computer science learners. This brought valuable insights for the main event in December - the International Computer Science Education Week.

### 4.2 Alice Game Jam

For 2015's Hour of Code ([7]) event, the game jam was aimed at engaging female teenagers and introducing them to programming in a playful way (Colley & Comber 2003). The theme of 'Alice in Wonderland' was selected because of the female protagonist, the story's bizarreness which provides a great basis for creative game ideas, and to celebrate the 150th anniversary of Lewis Carroll's novel. With this in mind, the idea for the Alice Game Jam event was born and conducted in cooperation with the Scratch team.

Based on the feedback of the previous game jam in October, the team provided a number of video tutorials. For each tutorial ([8]), a sample game was provided. Although the theme and the diversifiers where the same, the surprising topic was new: 'Time is running out!'. These can be found on https://share.catrob.at/pocketalice/.

A website was developed to support this event ([9]). Posters were sent to all of the schools involved in the NOLB project, with further advertising through NOLB partners and social media such as Facebook and Twitter.

Along with the general theme of Alice in Wonderland, our surprise topic was `Have I gone mad?'. Moreover learners could choose a maximum of four of the following diversifiers:

- Using sensors;
- Implementing at least two levels;
- Checking the learning content, e.g., through a quiz;
- Integration of collision detection;
- Using a foreign language.

By the end of the game jam, 200 Alice-themed programs (24% of them from female participants) were uploaded. The games comprised a wide range of fascinating and entertaining projects. All games were uploaded to pocketcode.org and can be found by searching for the hash tag #CodeEU.

Our findings indicate that slightly more than half of the submissions were created in small teams, thus identifying the potential of enabling skills such as sharing, team problem solving, cooperation, and collaboration. Thus, game jams in classrooms have the potential to support the development of learner's social and academic attitudes.

The data also identified that games were created across many school subjects. Game Jams can be adapted to support learning and teaching strategies across different disciplines and obviously do not need be restricted to computer science classes.

The analysis of our questionnaire data indicates that learners can be more motivated through game Jams and that learners who are less likely to create games are nevertheless more engaged in a Game Jam setting. Qualitative comments from learners indicated that older learners, ie learners over the age of 16 did not enjoy the theme of Alice finding it 'childish', while 100% of those 16 and under really enjoyed the theme feeling that it offered lots of possibilities. Learners used a range of diversifiers with sensors, collision detection and using a foreign language being the most popular. Thirty seven learners had used only one diversifier, the most popular for those using only one was using a foreign language.

---

[7] https://csedweek.org/learn
[8] https://share.catrob.at/pocketcode/gaming-tutorials
[9] http://www.alicegamejam.com/

## 5.1 BEST PRACTICE EXAMPLES

From the many submissions of the Alice Game Jam this section describes three examples. These programs not only followed the game jam theme but also reflect a range of disciplines such as Chemistry, Languages, and Maths. Additionally they consider diversifiers such as collision detection, the use of sensors, checking the learning content, and integrating the 'The Shape of a Game' ceremony framework.

Sick Alice: ID 5237 (see Figure 2)

| Subject | Language | Gender | Age |
|---------|----------|--------|-----|
| Chemistry | English | 2 x male | 17 |

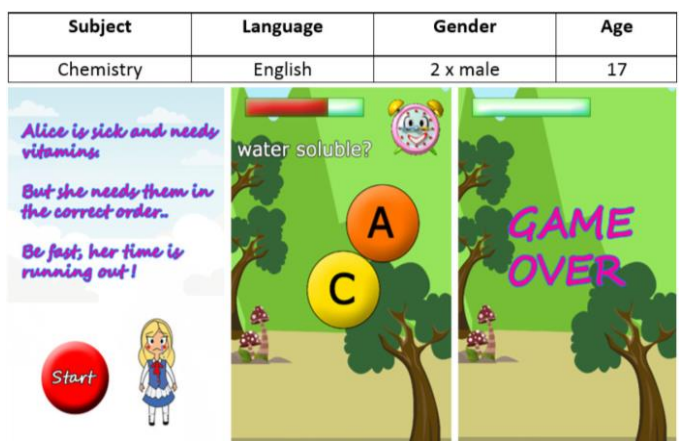

*Figure 2: Best practice example: Sick Alice.*

Description: Alice does not feel well, and it is your turn to help her get healthy again by tapping on the right vitamins. You have to differentiate between water-soluble and fat-soluble vitamins. Be aware: Time is running out.

Skater Alice: ID 5085 (see Figure 3)

| Subject | Language | Gender | Age |
|---------|----------|--------|-----|
| Languages | German and Turkish | 2 x female | 11/12 |

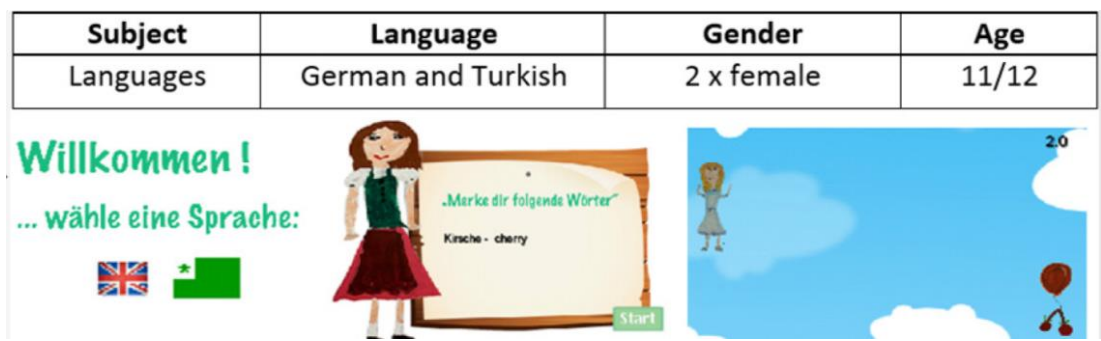

*Figure 3: Best practice example: Skater Alice.*

Description: This is a vocabulary game in landscape mode. Memorize the words. You can control Alice with the inclination sensors of the phone. Try to catch the objects from the list.

Concurso Alicia plantilla: ID 5238 (see Figure 4)

| Subject | Language | Team size | Gender | Age |
|---------|----------|-----------|--------|-----|
| Maths | Spanish | >3 | ? | 16 |

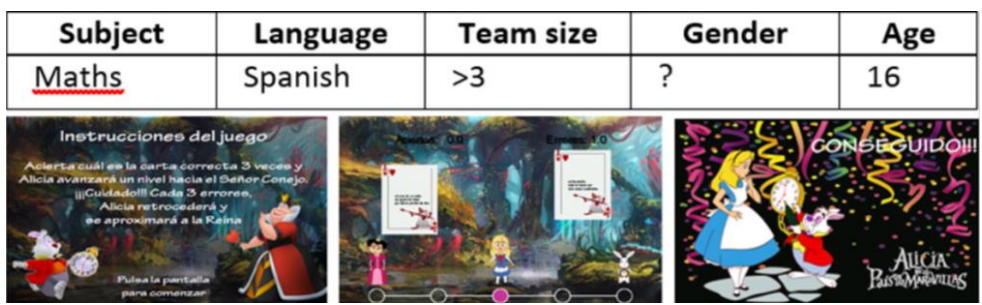

*Figure 4: Best practice example: Concurso Alicia.*

Description: Alice should try to verify math related assertions, e.g., 'the perimeter of a polygon is equal to the sum of its angles'. Every correct answer brings her closer to the rabbit and every wrong answer closer to the evil red queen.

# 6. CONCLUSION AND FUTURE WORK

This paper has explored the use of the mobile app, Pocket Code, which can be used on smartphones or tablets, being used in two game jams during the academic year 2015-16 with learners aged 11-18. A range of data has been collected and analysed. Our data evidences that learners can be more motivated through game jams and that learners who are less likely to create games are nevertheless more engaged in a game jam setting. Furthermore learner's various talents are nurtured by building and enriching personal and collaborative knowledge, and becoming part of a wider community with different social and cultural perspectives: this is a rich environment for learning which is still under-researched.

We have presented the NOLB Gaming Framework and the frameworks for 3 games from different disciplines: Chemistry, Languages, and Mathematics. One goal for the future will be to continue to design game jams specifically for schools and as new versions of Pocket Code become available. These will ensure a rich gaming experience for learners by providing helpful material during the jam.

The notion of game jams has provided a paradigm for creating both formal and informal learning experiences such as directed learning experience, problem-solving, hands-on projects, working collaboratively, and creative invention, within a learner-centred learning environment where children are creators of their own knowledge and learning material. These learning experiences have taken place in school, at home and in transit.

In considering what game jams can teach learners about design and programming in creative environments, it is clear that the playful setting of game jams helps learners understand and use programming languages, design standards for games, and also curriculum content within an inspiring and collaborative atmosphere. However, in the performed game jam events we must also consider the voluntariness of attending (school project/choose freely) and an imbalance between gender participation. There is scope for future game jams to further test different topics in engaging each gender. The experiences from these game jams will be utilised to further develop Pocket Code and the analytics that can be provided through Pocket Code and used by teachers in curriculum planning.

An output from this project will be use cases for online game jams in schools to identify how such events can be created by teachers and emerging pedagogy. The steps will include a) support for teachers using gaming in their classrooms for the first time, b) the creation and subscription phase, c) the coding and uploading phase, and d) the evaluation and presentation phase of the games. Materials to support teachers in using game jams will be a feature of the NOLB website referred to above.

The results of the performed jams will influence our planned game jam events in 2016 and 2017 and will help us to understand how game jams can be used for research and how to design them efficiently for educational purposes.

# 8. ADDITIONAL AUTHORS

Additional authors: Eugenio Gaeta (Life Supporting Technologies, Universidad Politecnica de Madrid email: eugenio.gaeta@lst.tfo.upm.es) and Jonathan Smith (GameCity and the UK National Videogame Arcade, email: jonathan@gamecity.org). We also acknowledge the contribution of the wider NOLB team at Nottingham Trent University: Professor David Brown, Jamie Tinney, Nick Shopland and Andy Burton, and the NOLB team at the National Videogame Arcade: Dominic Martinovs and Rachel Barrett.

BibTex entry:

@InProceedings{BOULTON2016EDU,
author = {Boulton, H. and Spieler B. and Petri, A. and
Schindler, C. and Slany W. and Beltràn, M.E.
title = {The Role of Game Jams in developing Informal
Learning of Computational Thinking: a cross-European
Case Study}
booktitle = {EDULEARN16 Proceedings},
isbn = {978-84-608-8860-4},
issn = {2340-1117},
doi = { 10.21125/edulearn.2016 },
url = {http://dx.doi.org/10.21125/edulearn.2016 },
publisher = {IATED},
location = {Barcelona, Spain},
month = {4-6 July, 2016},
year = {2016},
pages = {7034-7044}}